\documentclass[12pt]{iopart}

\usepackage{graphicx}
\usepackage{color}
\usepackage{amssymb}
\usepackage{placeins}

\def\bra#1{\langle #1|}
\def\ket#1{|#1\rangle}

\def\vp{\varphi}
\def\vr{\varrho}
\def\omegaeff{\omega_{{\rm eff}}}

\def\op#1{\hat{#1}}

\def\unit#1{\ \mathrm{#1}}
\def\sgn#1{\mathop{\rm sgn}}

\definecolor{orange}{rgb}{0.7,.35,0}

\begin{document}

\title{Tight-binding model for strongly modulated two-dimensional
superlattices}

\author{Karel V\'yborn\'y\dag\ddag\footnote[3]{To
whom correspondence should be addressed (vybornyk@fzu.cz)}\ 
and Ludv\'\i k Smr\v cka\dag }

\address{\dag\ Institute of Physics, Academy of Sciences of the Czech Republic,
Cukrovarnick\'a 10, 16253 Praha, Czech Republic,}

\address{\ddag\ I. Institute of Theoretical Physics, University
of Hamburg, Jungiusstr. 9, 20355 Hamburg, Germany}

\begin{abstract}
     Common models describing magnetotransport properties of
     periodically modulated two--dimensional systems often either
     directly start from a semiclassical approach or give results
     well conceivable within the semiclassical framework. Recently,
     magnetoresistance oscillations have been found on samples with
     strong unilateral modulation and short period ($d=15\unit{nm}$)
     which cannot be explained on a semiclassical level (magnetic
     breakdown \cite{deutschmann:02:2000}). We use a simple fully
     quantum mechanical model which gives us both magnetoresistance
     data nicely comparing to the experiments and a good intuitive
     insight into the effects taking place in the system.
\end{abstract}



Beginning with the pioneering work of Weiss {\it et
al.}\cite{weiss:01:1989} much effort was dedicated to magnetotransport
properties of periodically modulated two dimensional systems (2DES). A
wide palette of structures has been studied both experimentally and
theoretically: with periodic modulation in one direction or in both
directions, with modulation by either or both electric and
magnetic field, with various modulation strengths and concentrations of
electrons (see \cite{vyborny:06:2002} and references therein). Despite
the complexity of many such systems, the experimental data can usually be
interpreted within a semiclassical (SC) picture, based on constructing
classical trajectories of charge carrier and (if necessary) imposing a
quantization condition which reflects the formation of Landau levels in the
one--electron spectrum.

\begin{figure}
\begin{center}
\includegraphics[scale=0.3]{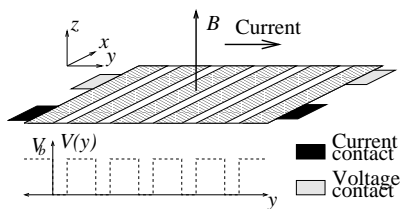}
\end{center}
\caption{Two--dimensional electron gas with periodic unilateral
modulation. Alternatively, the system may be conceived as an array of
wires coupled by tunnelling.}
\label{fig1_1}
\end{figure}

In this paper, we refer to a system where the SC prediction contradicts
the experimental finding. It is a strongly unilaterally modulated 2DES
with modulation period as short as $15\unit{nm}$ (see Section
\ref{secII} for criterion of short period); it may thus be conceived
as an array of weakly coupled quantum wires, see sketch in
Fig.~\ref{fig1_1}.  Experimentally, these conditions were achieved in
a GaAlAs/GaAs superlattice fabricated by cleaved edge overgrowth
technique first reported by Deutschmann {\it et al.}
\cite{deutschmann:02:2000}, see also more detailed description in
\cite{deutschmann:02:2001,deutschmann:2001}.

\begin{figure}
\begin{center}
\includegraphics[scale=0.3,angle=-90]{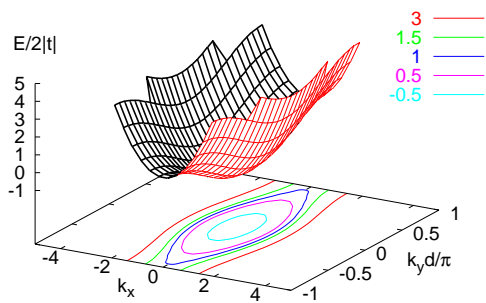}
\end{center}
\caption{Band structure in zero magnetic field: all modulation bands
(in $y$--direction) except for the lowest are discarded. Fermi
contours (related to real space trajectories) are closed for
$-2|t|<E_F<2|t|$ and open for $E_F>2|t|$.}
\label{fig2_1}
\end{figure}

\section{Model for Zero Magnetic Field: Semiclassics}

Since the potential in our system is separable, zero magnetic field
band structure is an effectively one--dimensional problem. Motion
along the wires ($x$ direction) is free and motion across the wires
($y$ direction) can be described by a Kronig--Penney model. For the
structure studied in \cite{deutschmann:02:2000}, we find that the
lowest band of the one--dimensional band structure in the $y$
direction is narrow ($4|t|\approx 3.8\unit{meV}$ wide), has nearly a
cosine form and is well separated from the higher bands ($\approx
60\unit{meV}$). Regarding the experimentally accessible concentrations
of electrons the Fermi level $E_F$ lies always well below the second
band (that means naturally also deep below $V_b$, the potential of
barriers in the superlattice) and thus, from now on, we will discard
all but the lowest modulation band. Note that this considerable
simplification is rendered by the shortness of the modulation period
(which introduces large modulation band gaps). The zero field band
structure is therefore (see Fig. \ref{fig2_1})
\begin{equation}\label{eq1_1}
  E(k_x,k_y)=\frac{\hbar^2k_x^2}{2m}-2|t|\cos k_yd\,.
\end{equation}

Let us now examine the SC model of a system with such a band
structure. The real space trajectory of an electron subject to a
perpendicular magnetic field $B$ will be the Fermi contour
$E_F=E(k_x,k_y)$ scaled by $\ell^2=\hbar/eB$ and rotated by
$90^\circ$. Evidently, this is a closed trajectory for $-2|t|<E_F<2|t|$
(resembling circular orbits of electrons moving in a plane,
i.e. 2D--like behaviour) and an open trajectory for $E>2|t|$
(corresponding to motion along the wires, i.e. 1D--like
behaviour). Shubnikov--de Haas oscillations of the magnetoresistance are
predicted for the former case (through the SC quantization condition:
real space trajectory must enclose integer multiple of magnetic flux
quanta $\Phi_0=h/e$, \cite{ashcroft:1976:p232}). In contrast,
the magnetoresistance is expected to be non-oscillatory in the latter case
(there is no quantization condition for open trajectories).

This model, however, does not always agree with the experiments
\cite{deutschmann:02:2000}: $1/B$--periodic oscillations were observed
even for $E_F>2|t|$ (see Fig. \ref{fig3_1}) being referred to as
magnetic breakdown. Since the type of periodicity is the same as for
$-2|t|<E_F<2|t|$ we can extend the SC model and claim that electrons
tunnel between the open trajectories and, thus, form loops for which the
usual quantization condition is to be fulfilled. This is however an
assumption strange to classical theories and we may ask how many such
{\it ad hoc} assumptions we need in order to reconcile theory and
experiment while pretending that electrons in such systems always obey
classical and not quantum mechanical laws.

As it was mentioned above this situation occurs owing to the strong
modulation (i.e. width of the lowest modulation band is as low as the
Fermi level, $E_F\approx 4|t|$) and the shortness of the modulation period
($d^2\lesssim 3h^2/(2|t|m)$).

\section{Quantum Mechanical Model}
\label{secII}

The system is described by the Hamiltonian
\begin{equation}\label{eq3_3}
  H=\frac{1}{2m}(p_x+eBy)^2+\frac{1}{2m}p_y^2+V(y)
\end{equation}
in the Landau gauge $\vec{A}=(By,0,0)$ (we set $e=|e|$). The
restriction to the lowest modulation band is equivalent to the
tight--binding ansatz (used also by Wulf {\it et al.} \cite{wulf:09:1992})
$$
  \Psi(x,y)=\frac{1}{\sqrt{2\pi}}\exp(ikx)\psi_{k,n}(y)
  =\frac{1}{\sqrt{2\pi}}\exp(ikx)\sum_j a_j(k)\vp(y-jd)\,,
$$
$$
    \bra{\vp_i} H_y\ket{\vp_j}=-|t|\delta_{i,j\pm 1}
$$
where $\vp(y-jd)$ or $\ket{\vp_j}$ denotes the ground state (more
precisely Wannier state) in the $j$-th well of the modulation potential
and $n$ is Landau band index.
Moreover, we assume that $t$ does not change with magnetic field. This
is plausible unless the magnetic field is extremely strong ($\ell^2k_F\ll
d$, $k_F$ is the Fermi wavevector; in such a case a 2DES is formed
inside one quantum wire). 

The Hamiltonian can be now written as a matrix
\begin{eqnarray}\label{eq3_6}
  H_{jl}(k)&=&|t|\left[\alpha^2\big((k/K)+j\big)^2\delta_{j,l}-
         \delta_{j,l\pm  1}\right]\,,\\ \nonumber
         &&\alpha^2=\frac{e^2B^2}{m}\cdot\frac{d^2}{2|t|}=
	 \left(\frac{\hbar\omegaeff}{2|t|}\right)^2\,,\qquad
	 K=d\frac{eB}{\hbar}\,
\end{eqnarray}
which depends (up to the scaling of $k$ and energy) 
on a single dimensionless parameter~$\alpha$. 

\begin{figure}

\begin{center}
\unitlength1.5mm
\begin{tabular}{ccc}
\includegraphics[scale=0.2475]{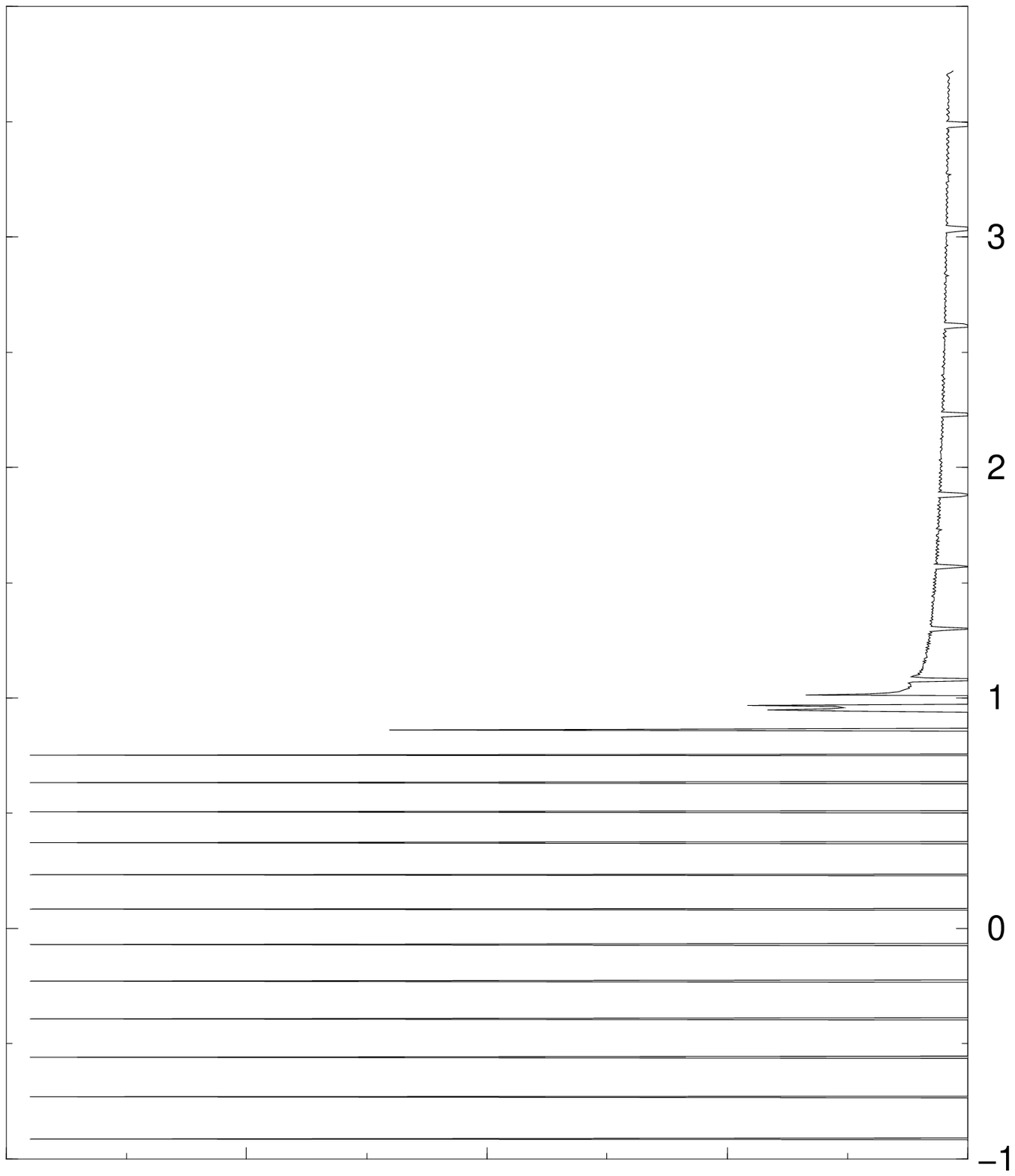}%
\hskip-3.75cm\raise.48cm\hbox{\includegraphics[width=1.65cm,height=3.3cm,angle=90]{cos1.epsi}}\put(-16,0.8){{\small DOS, $\xi$ [a.u.]}}
&
\raise-1.5mm\hbox{\includegraphics[scale=0.2475]{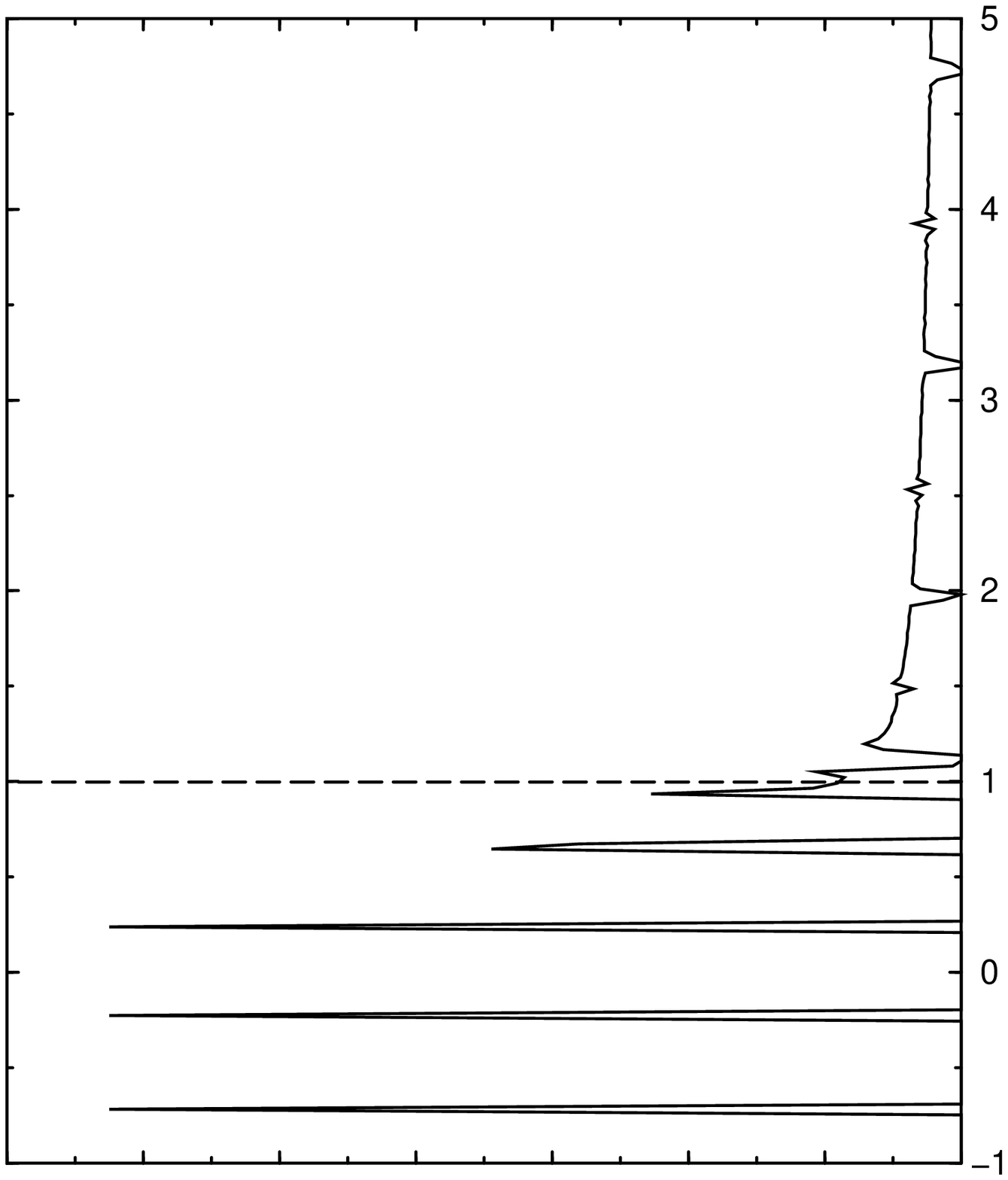}%
\hskip-3.6cm\raise.6cm\hbox{\includegraphics[width=1.395cm,height=3cm,angle=90]{cos2.epsi}}\put(-16,1.7){{\small DOS, $\xi$ [a.u.]}}}
&
\raise-3mm\hbox{\includegraphics[scale=0.2475]{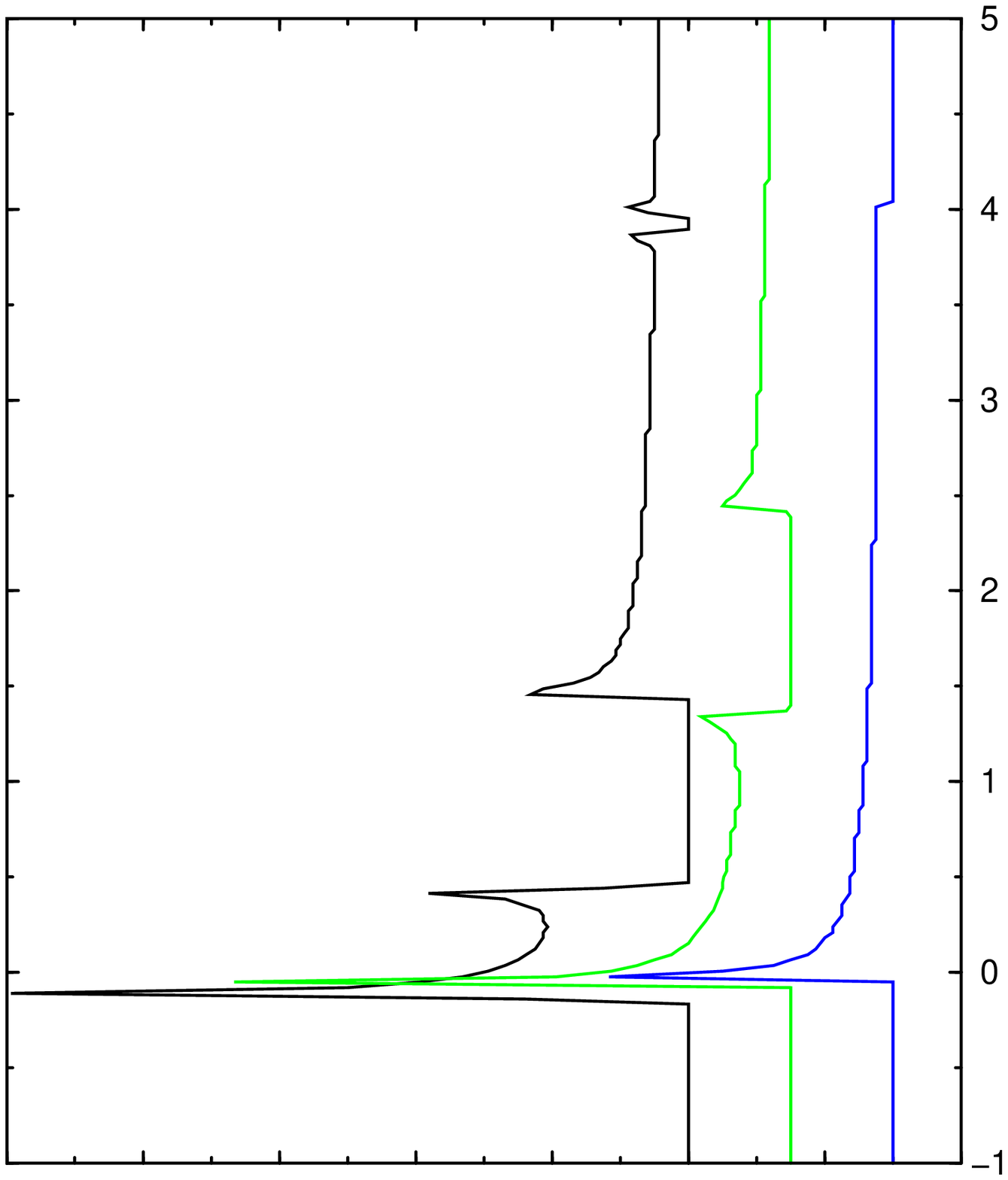}%
\hskip-3.6cm\raise.795cm\hbox{\includegraphics[width=1.395cm,height=3cm,angle=90]{cos3.epsi}}\put(-16,2.7){{\small DOS, $\xi$ [a.u.]}\put(7.4,15){\rotatebox{90}{{\small $E/2|t|$}}}}}
\\[2mm]
$\alpha\ll 1$ & $\alpha\approx 1$ & $\alpha> 1$ \\
weak field & intermediate field & strong field \\
\end{tabular}
\end{center}
\caption{Density of states at different magnetic fields and
corresponding potential for the fictitious particle. At low magnetic
field ($\alpha\ll 1$) there are almost sharp Landau levels for
$-2|t|<E<2|t|$ and an almost 1D--like DOS ($\propto 1/\sqrt{E}$) for
$E>2|t|$: this corresponds to the semiclassical model. At higher
magnetic fields ($\alpha\approx 1$) the Landau levels broaden into
Landau bands (an effect of the modulation) and gaps in the
1D--like region become more pronounced. Finally for $\alpha>1$ there
remains only one broad Landau band in the $-2|t|<E<2|t|$ region and
the gaps still grow (for our system parameters the DOS displayed in
the graph on the right correspond left to right to $B=5$, $7$ and $11$
Tesla).  }
\label{fig2_2}
\end{figure}

If the system is infinite in the $y$ direction, the spectrum is
$K$--periodic in $k$ (due to invariance to magnetic translations) and
its spectrum coincides with the one of a {\em fictitious} 1D particle in a
periodic cosine potential (Mathieu equation) \cite{ICPSpaper:note1}
\begin{equation}\label{eq3_5}
  -\frac{\hbar^2}{2m}\chi''(\xi)-\chi(\xi)2|t|\cos K\xi=E\chi(\xi)\,.
\end{equation}
Note, that the cosine potential changes with magnetic field (via $K$).
This allows for an easy notion of what the spectrum looks like for
different values of $\alpha$ (see Fig. \ref{fig2_2}). For instance, if
the period of the cosine potential is large (it means small $K$, more
precisely $\alpha\ll 1$), each of the periods contains a broad
potential well: near the bottom it may be approximated by a quadratic
potential and we thus obtain almost equidistantly spaced levels. Since
these cosine wells are only weakly coupled to each other (barriers
between them are thick), the band structure will comprise of narrow
bands (Fig \ref{fig2_2} left). On contrary, states with an energy
$E>2|t|$ high above the top of the cosine potential (but still under
the second modulation band, i.e. deep under $V_b$, see
Fig. \ref{fig1_1}) will percept this potential only as a perturbation:
the spectrum will be almost as of a free 1D particle (parabolic) and
with small gaps opening at $k=\pm\frac{1}{2}K$ due to the underlying
potential perturbation.

\begin{figure}
\begin{center}
\includegraphics[scale=0.3,angle=-90]{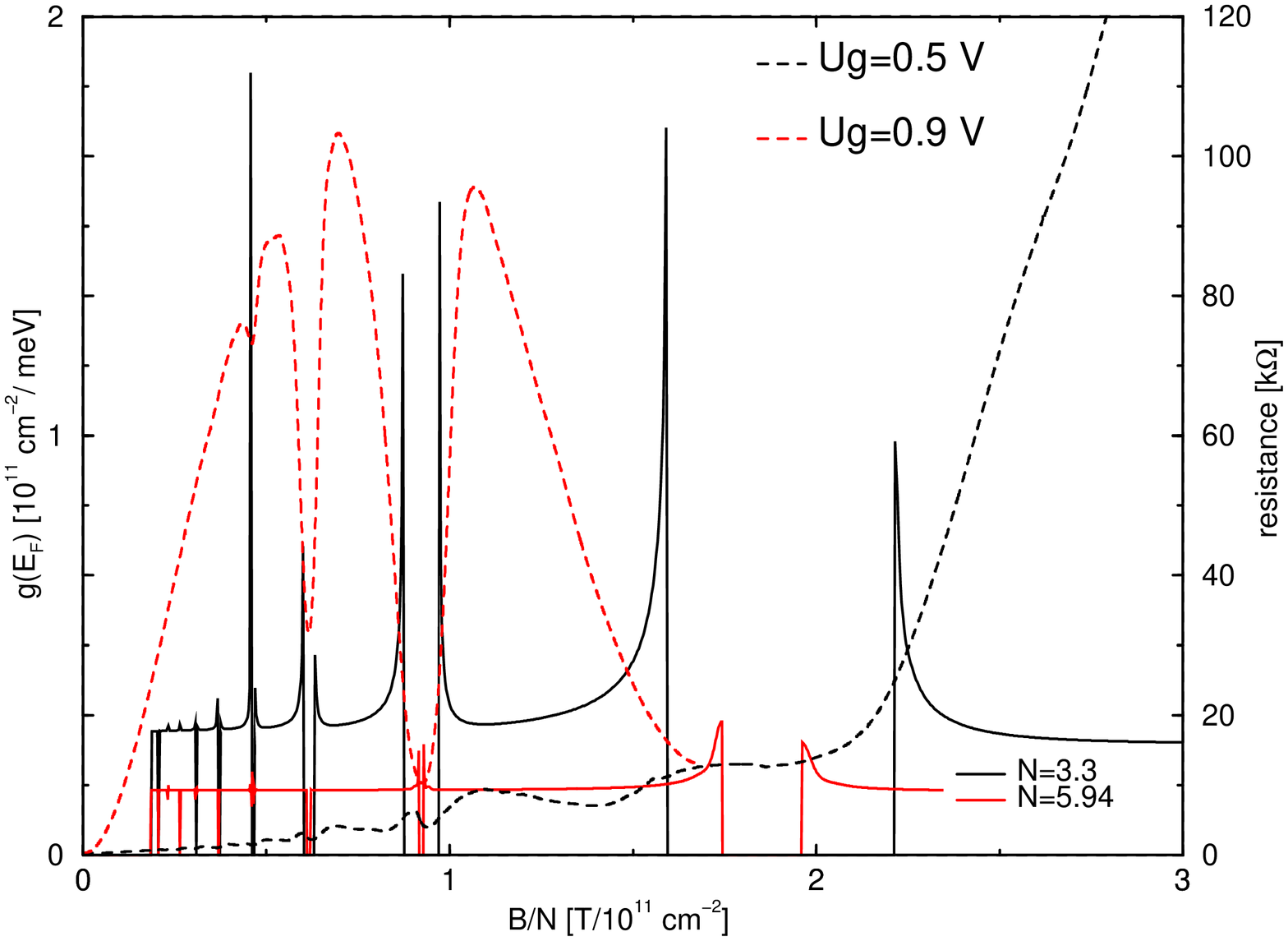}
\includegraphics[scale=0.3,angle=-90]{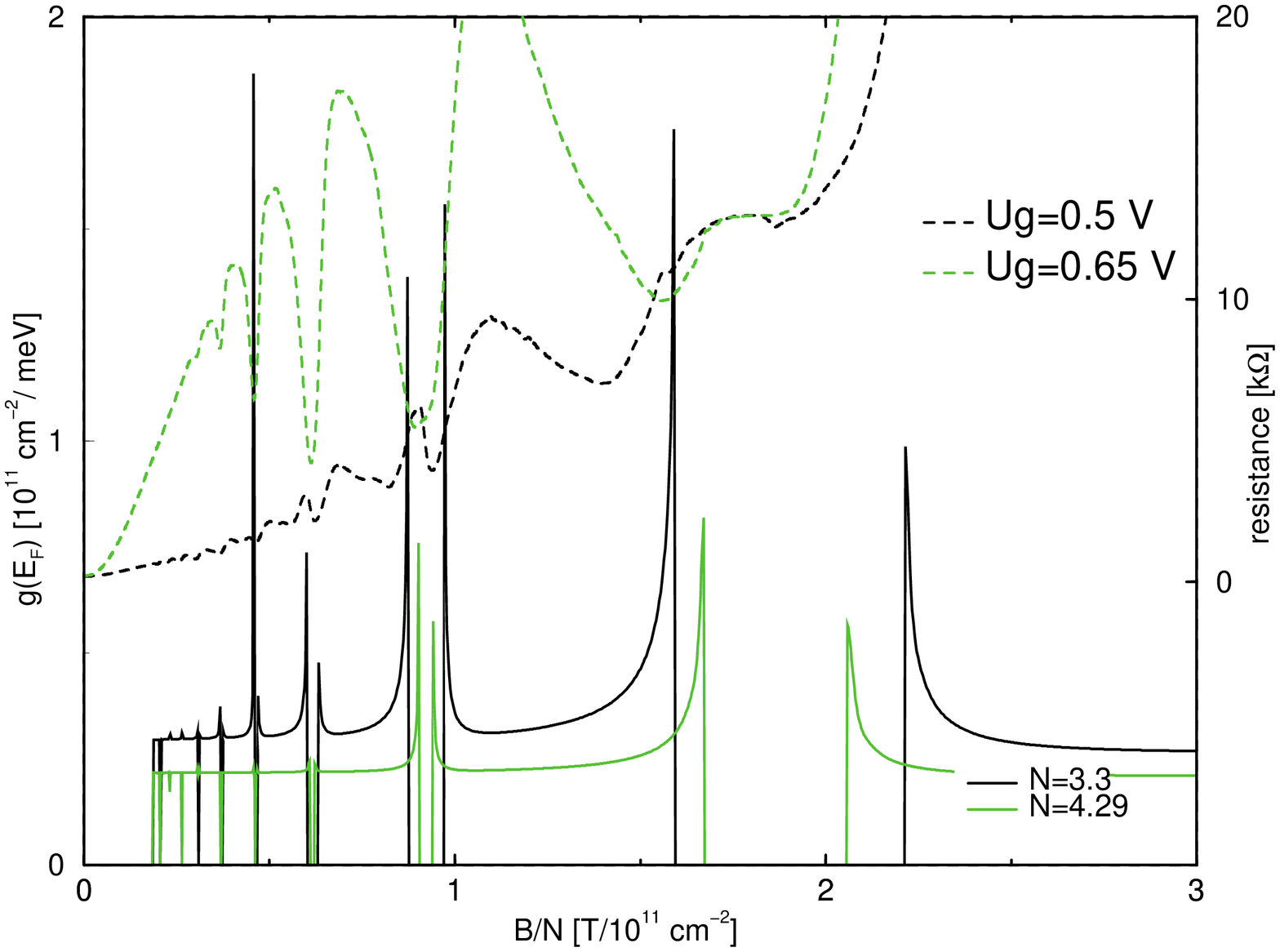}
\end{center}
\caption{Density of states (solid lines) compared to the experimental
magnetoresistance (dashed lines; reprinted with kind permission of
R.~A.~Deutschmann): gaps in the DOS agree with the extrema of the
magnetoresistance. Gate voltages are given for the experimental curves
and corresponding 2D electron concentrations in
$10^{-11}\unit{cm^{-2}}$ are given for the DOS. All the displayed
curves correspond to $E_F>2|t|$.}
\label{fig2_3}
\end{figure}

Plotting the density of states (DOS) at the Fermi level as a function
of magnetic field we can see that our model reproduces very precisely
the magnetoresistance oscillations for various electron concentrations
(Fig. \ref{fig2_3}).

\section{Transport calculations}
\label{secIII}

Based on the Kubo formula (linear response to applied bias)
we computed the conductivity tensor components (see \cite{vyborny:06:2002})
at zero temperature (for brevity we write $E$ for the Fermi level)
\begin{eqnarray*}
   \sigma_{xx}(E) &=&
  \frac{2}{\pi\Gamma
  d}\cdot\frac{e^2}{h}\cdot\frac{\sgn\big(g(E)\big)}{g(E)}+\\
  &&+  \frac{4\pi\Gamma}{d}\cdot
  (\hbar\omega)^2\cdot\frac{e^2}{h}\cdot
  g(E)\sum_{n'\not=n} 
  \left(\frac{\bra{\psi(k,n')} \op y \ket{\psi(k,n)}}{E(k,n')-
  E(k,n)}\right)^2\\
  \sigma_{yy}(E)&=&\frac{4\pi \Gamma}{d}\cdot\frac{e^2}{h} 
  g(E)\sum_{n'\not=n} 
  \Big(\bra{\psi(k,n')} \op y \ket{\psi(k,n)}\Big)^2\\
  \sigma_{xy}(E)&=&e\frac{\partial N(E)}{\partial B}+
  \frac{4\pi\hbar\omega}{d}\cdot\frac{e^2}{h}g(E)
  \sum_{n'\not= n}
  \left(\bra{\psi(k,n)}\op y \ket{\psi(k,n')}\right)^2\,,
\end{eqnarray*}
including the factor two for spin. Here $\omega=eB/m$, $N(E)$ is
number of states with energy less than $E$, $\op y$ is the
$y$--coordinate operator and $\sgn\big(g(E)\big)/g(E)$ is to be
understood as 'zero for $g(E)=0$ and $1/g(E)$ else'. We treated the
impurity scattering within the c--number approximation for
self--energy ($\Gamma$ denotes its imaginary part); $\Gamma$ has then
the meaning of an inverse relaxation time, $\Gamma=\hbar/\tau$. 

The first term of the $xx$--component of conductivity can be
attributed to electrons moving along open orbits. If the wires in the
superlattice were decoupled, the DOS would be the same as of a 1D
electron gas ($g(E)\propto 1/\sqrt{E}$) and the first term in
$\sigma_{xx}$ would be simply proportional to $\sqrt{E}$, i.e. the
electron velocity. This term is proportional to the relaxation time
(as may be expected).  In contrast, $\sigma_{yy}$ (as well as the
second term of $\sigma_{xx}$) is inversely proportional to the
relaxation time: this means that conduction across the wires is due to
{\em inter}--Landau--band scattering (sumation over $n'\not=n$)
introduced by impurities.

The first term in $\sigma_{xy}$ introduces Hall plateaus of quantized
conductivity when $E$ lies in a gap; it vanishes in the classical
limit. The second term does not depend on the relaxation time. It is
proportional to $B$ (appearing in $\omega$) and resembles thus the
classical Hall conductivity.

\begin{figure}
\begin{center}
\includegraphics[scale=0.5,angle=-90]{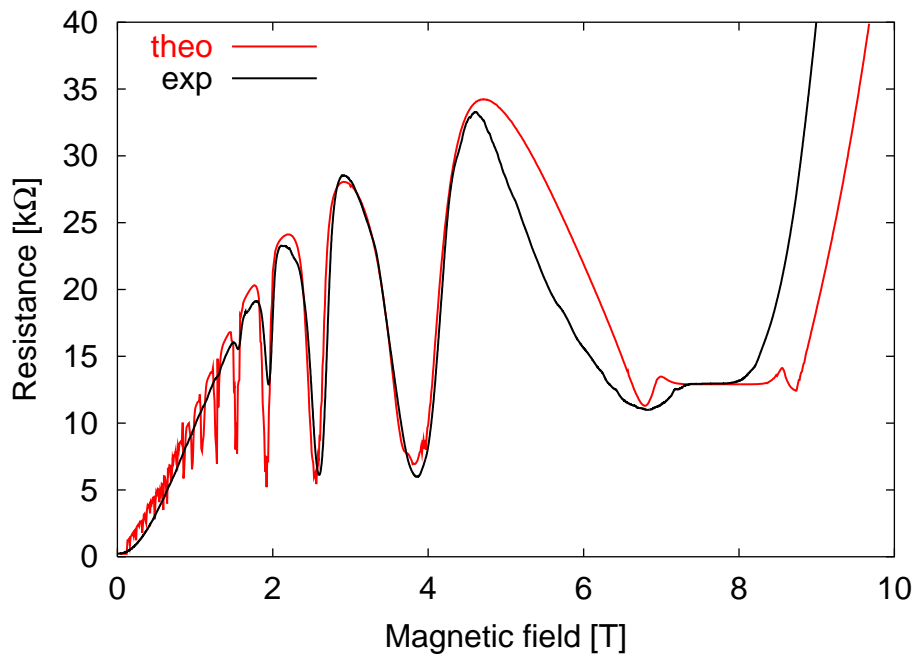}
\end{center}
\caption{Theory prediction and experimental magnetoresistance
(reprinted with kind permission of R.~A.~Deutschmann). According to
the semiclassical theory there should be no oscillations
($E_F>2|t|$). For the imaginary part of the self--energy (inverse
relaxation time) the phenomenological ansatz $\Gamma(E,B)\propto \sqrt{B}$
was used (according to \cite{manolescu:08:1995}).}
\label{fig3_1}
\end{figure}

To be able to discuss the experimental data we have to address
the relation between the computed conductivity components and measured
resistance at last. Due to technological reasons resistivity
components $\vr_{yy}$ and $\vr_{xy}$ could not be measured 
separately. We expect that the measured resistance is a mixture of
these two quantities, $c\vr_{yy}+\vr_{xy}$ and this allows us to fit
the experimental curves quite well with a single fitting parameter
(Fig. \ref{fig3_1}) for Fermi energies $E_F>2|t|$. The agreement is
worse but still at least qualitative for $E_F\approx 2|t|$.

\section{Conclusion}

We investigated a two--dimensional electron system for which the usual
semiclassical approach fails to predict correct magnetoresistance
behaviour. The wide gap between the lowest and second lowest
modulation band was the crucial prerequisite for this to happen which
in turn depends on the unusually short period of the modulation
potential.  The observed $1/B$--periodic magnetoresistance
oscillations may be understood as a consequence of electron tunnelling
between semiclassical orbits.  We presented a simple quantum
mechanical model based on a tight binding approximation which can
account very well for these oscillations.  Very simplistic transport
calculations are in nearly quantitative agreement with the experiment
for a wide range of electron concentrations. Outside this range
qualitative agreement is still retained.

One of the authors (KV) would like to acknowledge discussions with
Daniela Pfannkuche and Alexander Chudnovskiy.  This work has been
partly supported by the Grant Agency of the Czech Republic under
Grants No. 202/01/0754 and No. 202/03/0431.

\section{References}

\bibliographystyle{prsty}
\bibliography{bib_wiresICPS}

\begin{thebibliography}{1}

\bibitem{deutschmann:02:2000}
R.~A. Deutschmann {\it et~al.}, Physica E {\bf 6},  561  (2000).

\bibitem{weiss:01:1989}
D. Weiss, K. von Klitzing, K. Ploog, and G. Weinmann, Europhys. Lett. {\bf 8},
  179  (1989).

\bibitem{vyborny:06:2002}
K. {V\'yborn\'y}, L. {Smr\v cka}, and R.~A. Deutschmann, {cond-mat/0206212}  .

\bibitem{deutschmann:02:2001}
R.~A. Deutschmann {\it et~al.}, Phys. Rev. Lett. {\bf 86},  1857  (2001).

\bibitem{deutschmann:2001}
R.~A. Deutschmann, {\em Two dimensional electron systems in atomically precise
  periodic potentials}, {\em Selected topics of Semiconductor Physics and
  Technology, {Vol. 42}} (University of Technology Munich, Munich, 2001),
  {PhD.} thesis, ISBN 3-932749-42-1.

\bibitem{ashcroft:1976:p232}
N. Ashcroft and N. Mermin, {\em Solid State Physics} (Saunders College
  Publishing, Orlando, 1976), p.~232.

\bibitem{wulf:09:1992}
U. Wulf, {J. Ku\v cera}, and A. MacDonald, Phys. Rev. B. {\bf 47},  1675
  (1992).

\bibitem{ICPSpaper:note1}
To see the equivalence between Eq. \ref{eq3_6} and Eq. \ref{eq3_5}, substitute
  $\chi(\xi)=e^{ik\xi}\sum_n a_n e^{inK\xi}$ into Eq. \ref{eq3_5} in accordance
  with Bloch's theorem. Then $H_{jl}(k)$ turns out to be the matrix of the
  Hamiltonian in Eq. \ref{eq3_5} in the basis $e^{ik\xi}e^{inK\xi}$, $n=0,\pm
  1, \pm 2,\ldots$.

\bibitem{manolescu:08:1995}
A. Manolescu {\it et~al.}, Surf. Sci. {\bf 361/362},  513  (1996).

\end{thebibliography}

\end{document}